\documentclass[10pt,journal,compsoc]{IEEEtran}

\usepackage{amssymb}
\usepackage{amsmath}
\usepackage{graphicx}
\usepackage{wrapfig}
\usepackage[labelfont=bf]{caption}
\usepackage{float}
\usepackage{algorithm}
\usepackage[noend]{algpseudocode}
\usepackage{tabu}
\usepackage{array}
\usepackage{adjustbox}
\usepackage{multirow}
\usepackage[numbers]{natbib}
\usepackage[bottom]{footmisc}
\usepackage[usenames, dvipsnames]{color}
\newcolumntype{N}{>{\centering\arraybackslash}m{4em}}
\newcolumntype{R}{>{\centering\arraybackslash}m{8em}}
\newcolumntype{W}{>{\centering\arraybackslash}m{8em}}
\newcolumntype{Z}{>{\centering\arraybackslash}m{15em}}

\begin{document}

\title{Geometry Processing of Conventionally Produced Mouse Brain Slice Images}

\author{Nitin~Agarwal, Xiangmin~Xu and~M~Gopi
\IEEEcompsocitemizethanks{\IEEEcompsocthanksitem Nitin Agarwal and M.Gopi are in Computer Science Department of University of California, Irvine. E-mail: $\left\lbrace agarwal,gopi\right\rbrace$@ics.uci.edu.
\IEEEcompsocthanksitem Xiangmin Xu is in Department of Anatomy and Neurobiology at University of California, Irvine. Email: xiangmin.xu@uci.edu.}%
}

\setcounter{figure}{-2}
\IEEEtitleabstractindextext{
\begin{center}
\centering
\includegraphics[width=0.9\textwidth,height=5.2cm]{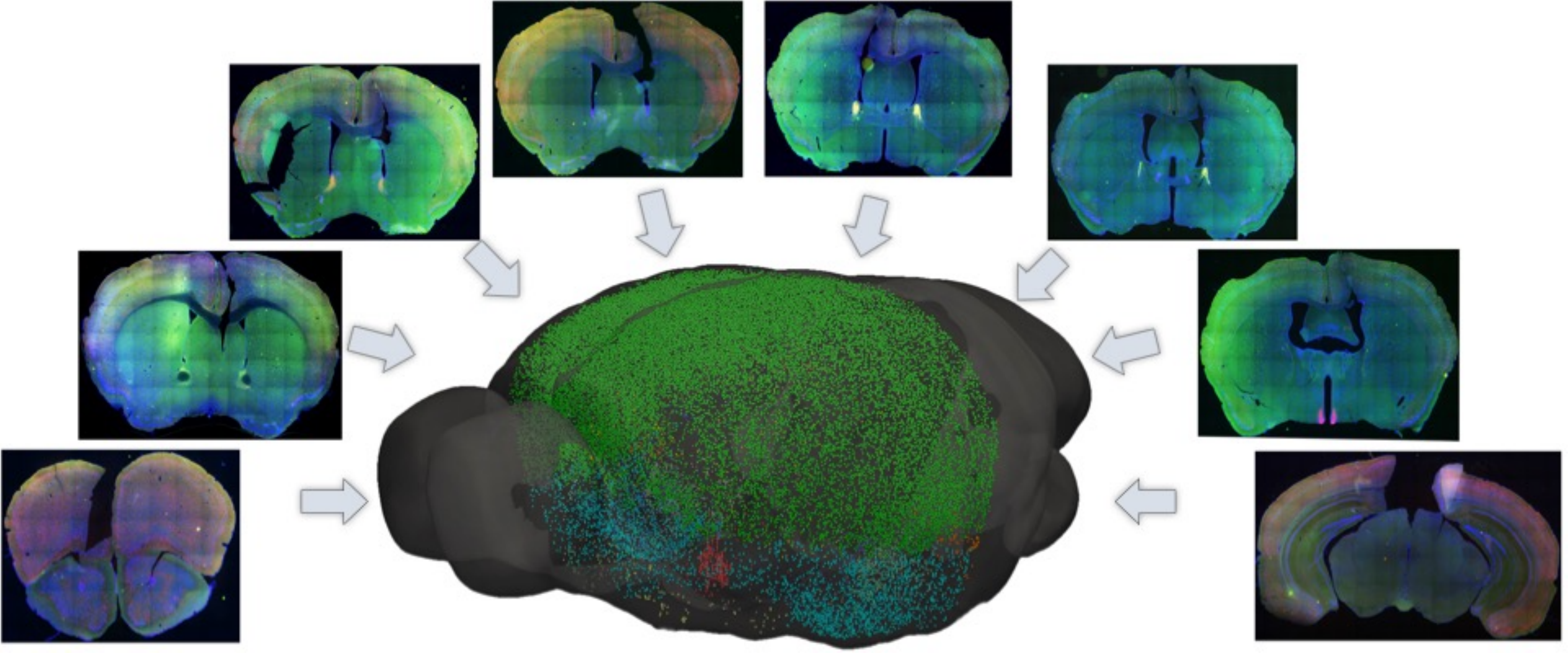}
\captionof{figure}{We automatically align conventionally produced mouse brain slices which may contain histological artifacts to an annotated 3D common reference space. This allows us to count the number of neurons in each anatomical region of the mouse brain.  Segmented neurons from 51 microscopic slices of a mouse brain are colored by their anatomical region and visualized inside a 3D virtual mouse brain model.}
\label{fig:teaser}
\end{center} 

\begin{abstract}
Brain mapping research in most neuroanatomical laboratories relies on conventional processing techniques, which often introduce histological artifacts such as tissue tears and tissue loss. In this paper we present techniques and algorithms for automatic registration and 3D reconstruction of conventionally produced mouse brain slices in a standardized atlas space. This is achieved first by constructing a virtual 3D mouse brain model from annotated slices of Allen Reference Atlas (ARA). Virtual re-slicing of the reconstructed model generates ARA-based slice images corresponding to the microscopic images of histological brain sections. These image pairs are aligned using a geometric approach through contour images. Histological artifacts in the microscopic images are detected and removed using Constrained Delaunay Triangulation before performing global alignment. Finally, non-linear registration is performed by solving Laplace's equation with Dirichlet boundary conditions. Our methods provide significant improvements over previously reported registration techniques for the tested slices in 3D space, especially on slices with significant histological artifacts. Further, as an application we count the number of neurons in various anatomical regions using a dataset of 51 microscopic slices from a single mouse brain. This work represents a significant contribution to this subfield of neuroscience as it provides tools to neuroanatomist for analyzing and processing histological data. 

\end{abstract}


}

\maketitle

\IEEEdisplaynontitleabstractindextext

\vspace{-0.1 cm}
\IEEEraisesectionheading{\section{Introduction}
\label{sec:intro}}
\IEEEPARstart{U}{nderstanding} the brain connectome or the wiring diagram of the brain is essential to understand how the brain circuits work \cite{van2012aa,Lichtman:2011aa}. However, obtaining the wiring diagram of the human brain is extremely difficult as it is large and contains billions of neurons forming complex interconnecting networks. Obtaining the connectome of even a simple roundworm such as C.elegan, which consists of only 302 neurons took many years \cite{White:1986aa}. Only recently, with the advances in both computing power and optical imaging techniques, it has now become feasible to obtain the connectome of more complex brains. A salient example of this is the ongoing efforts in mapping the connections in Drosophila's brain which has nearly 100,000 neurons \cite{Chiang:2011aa}. Over the past decade, neuroscience researchers have started studying the mouse brains due to their physiological and genetic similarity to humans, the ease with which their genomes can be manipulated, and the ability to train mice to perform behavioural tasks relevant to human cognitive processes. 

There are two steps in processing mouse brains. The first step comprises of sample preparation, imaging, and collection of histological slice data, while the second step consists of analyzing this histological data for measurement and quantification of labeled neurons, studying gene expression patterns, connectome exploration, etc. In most neuroanatomical laboratories, both these steps are largely performed manually \cite{Agarwal:2016aa,Sun:2014aa}. Manual sample preparation, although offers great flexibility especially in restaining of slices, slicing at arbitrary intervals etc., introduces many slice-specific histological artifacts such as tissue tears, folds, and missing regions. The second step, manual analysis of histological slices, is tedious, incomplete, and introduces various subjective errors. There is very little advantage to do the analysis, measurement, and visualization of histological slices manually. 

\begin{figure}[!b]
\centering
\includegraphics[width=0.49\textwidth,height=2.3cm]{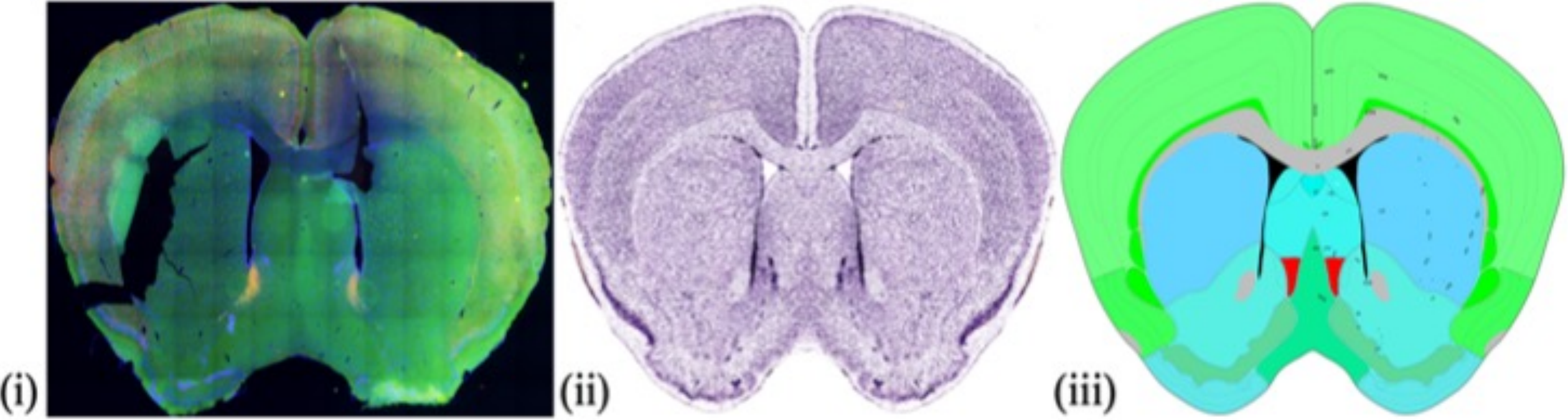}
\caption{\textbf{Example Images illustrating the challenges in histological alignment:} (i) Conventionally produced microscopic slice image with histological artifacts. (ii) Corresponding Nissl intensity slice and (iii) annotated atlas slice from Allen Reference Atlas. There exists no straight forward intensity relationship between the microscopic slice and the Nissl intensity slice.}
\label{fig:altasSlices}
\end{figure}

It is advantageous to allow manual sample preparation, if required, and automate the second step, namely the post processing of histological slice data. However, the artifacts introduced during manual sample preparation makes many post-processing operations such as automatic alignment and 3D reconstruction extremely challenging \cite{Yushkevich:2006aa, Agarwal:2016ab}. Another challenge in processing these conventionally produced slices is that a variety of sample preparation and staining procedures like In-Situ Hybridization (ISH), histology, etc., result in brain slices having different intensity profiles making comparisons with the reference atlas images extremely difficult (Fig. \ref{fig:altasSlices}). In this paper, we present algorithms and techniques to address this challenging task of automating the post-processing of mouse brain slices including those that are produced by conventional techniques.

In order to understand the wiring diagram of a mouse brain, it is crucial to visualize and explore the connectome data in a standardized brain space or a reference atlas. There exists many mouse brain reference atlases, each constructed using different procedures \cite{Paxinos:2004aa,Dong:2008aa, Valverde:1998aa}. Among these, the ARA \cite{Dong:2008aa} has been widely used in neuroanatomy laboratories in the world. ARA is being continuously updated and so far it has delineated approximately 738 mouse brain anatomical regions. ARA consists of two reference atlases (coronal and saggital reference atlas) created by slicing the mouse brain in different directions. The coronal reference atlas consists of 132 sections evenly spaced at 100 $\mu$m whereas the saggital reference altas consists of 21 sections spaced at 200 $\mu$m. Each of these reference atlases further consists of a stack of Nissl intensity slices and a stack of annotated contour slices hand drawn by experts as shown in Figure \ref{fig:altasSlices}.

In this paper, we report our techniques and algorithms to register conventionally produced microscopic mouse brain slices to a 3D annotated reference atlas space constructed from the ARA slices. A registration such as this has numerous advantages. First, it allows transfer of annotations from ARA onto the microscopic slices thereby facilitating region based neuron counting, analysis of common gene expression patterns etc. Second, inter-subject comparisons can be easily performed by registering multiple mouse brains to this 3D reference atlas space. Third, it enables virtual reslicing of the 3D reference atlas space creating new slices with annotations transferred from the ARA. Fourth, it supports 3D visualization and analysis of neuronal projections which often span the entire central nervous system providing functional connections between anatomically distant regions. 

\begin{figure}[!t]
\centering
\includegraphics[width=0.49\textwidth,height=6cm]{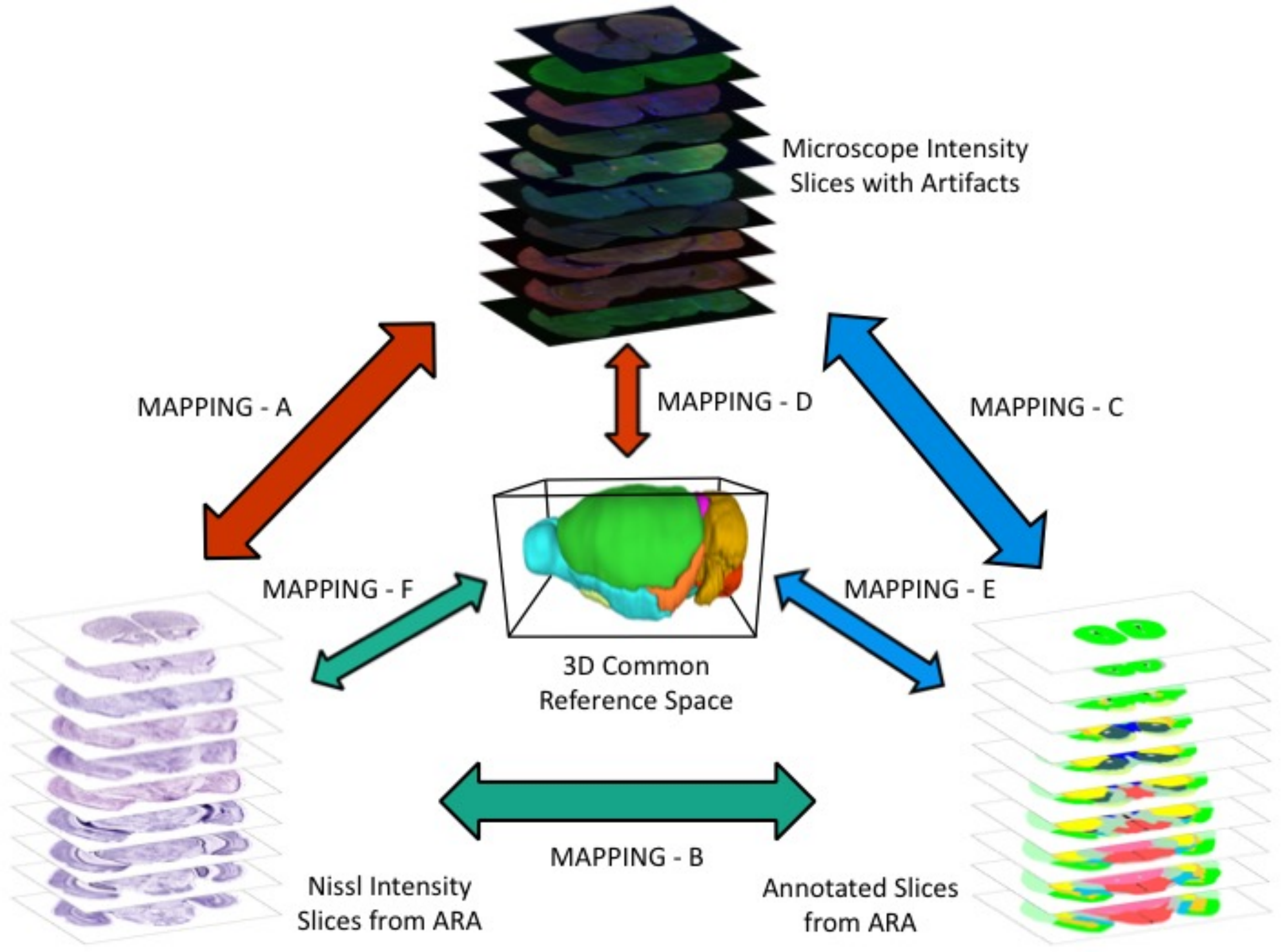}
\caption{\textbf{Construction of annotated 3D common reference space:} This can be accomplished through 3D reconstruction and alignment of Microscopic slices (D) and either Annotated atlas slices (E), or Nissl intensity slices (F). Another approach could be 2D registration of Microscopic slices with either the Nissl intensity slices (A) or the Annotated atlas slices (C) followed by 3D reconstruction of any one of the three spaces. Note that mapping B between Nissl intensity and annotated atlas slices can be deduced from ARA maps.}
\label{fig:pathways}
\end{figure}

\begin{figure*}[!ht]
\centering
\includegraphics[width=0.97\textwidth,height=7.2cm]{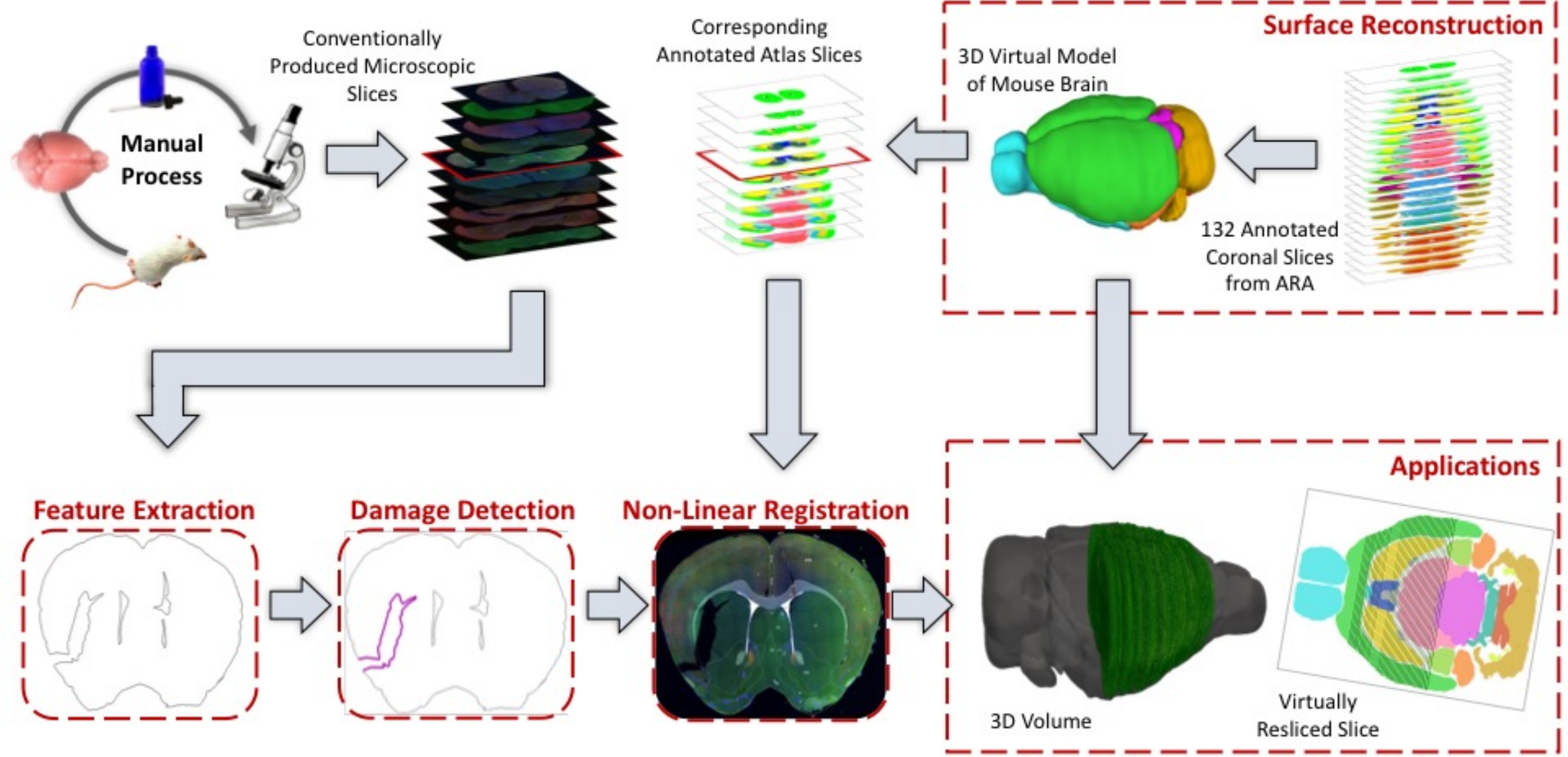}
\caption{\textbf{Illustration of our system pipeline:} We first reconstruct a 3D virtual mouse brain model from 132 coronal annotated slices from Allen Reference Atlas. This 3D model is then virtually resliced in the same direction and slicing interval that was used to slice the actual mouse brain so that each microscopic slice image has a corresponding annotated atlas slice image. These image pairs are then aligned using feature extraction, damage region detection and non-linear registration. Using the registered microscopic slices and the virtual mouse brain model, we can now perform 3D volume reconstruction of the original mouse brain. We can also transfer annotations from the reconstructed model onto any virtually resliced slice. Such applications provide tools to assist neuroanatomist in studying mouse brains.}
\label{fig:pipeline}
\end{figure*}

Construction of an annotated 3D common reference space in which the microscopic slices are aligned with the reference atlas can be broadly accomplished using either of the two approaches as shown in Figure \ref{fig:pathways}. Since the ARA consists of two reference spaces - the Nissl intensity slices and the annotated slices, one approach could be to first perform 3D reconstruction of the microscopic intensity slices (Mapping D) and either of the two reference spaces (Mapping F or Mapping E) and then perform 3D alignment between the two reconstructed 3D models. This is challenging because most previous works for 3D volume reconstruction from intensity slices assume the slices to have little or no artifacts \cite{Ju:2006aa,Kovacevic:2005aa,Nikou:2003aa} and hence can handle Mapping F but not Mapping D, especially when the slices have histological artifacts. Furthermore, Mapping E is also difficult as annotated ARA slices have large inter-slice distance (100 $\mu$m), whereas most previous methods require the inter-slice distance to be very small (approx 25 $\mu$m) to ensure topological correctness \cite{Ju:2005aa,Das:2005aa,Liu:2008aa}. Alternatively, to avoid topological issues during reconstruction, 3D surface reconstruction of different regions of the brain has to be performed which may later be converted to volume representation. Another possible approach for construction of the annotated 3D common reference space could be to first perform 2D alignment between microscopic slices and either of the two reference spaces (Mapping A or Mapping C) followed by 3D reconstruction of any of the three reference spaces (This is possible since Mapping B between the Nissl intensity slices and the annotated atlas slices is already known). The problem with this approach is that microscopic slices produced from conventional techniques are riddled with histological artifacts making this \emph{inter-stack} registration problem extremely difficult. Further, intensity based registration approaches (Mapping A) require the microscopic slices to have a similar intensity profile as the Nissl intensity slices, a requirement that is very difficult to satisfy.  Hence, we propose to register the microscopic image stack to an annotated 3D reference space (Mapping D) in two steps (blue): \emph{Mapping C: }2D registration of microscopic slices with the annotated atlas slices, and \emph{Mapping E: }3D surface reconstruction from annotated slices of ARA. During this process we make the following main contributions.

\noindent
\textbf{Main Contributions}:
\begin{itemize}
\item \emph{3D surface reconstruction:} We perform 3D surface reconstruction from annotated slices of ARA which have large inter-slice distance to generate a smooth virtual mouse brain model with correct topology and no intersections. During this process we reconstruct 20 major anatomical regions, in which each region is a manifold. The final high-resolution mesh comprises of approximately $307K$ vertices and $615K$ faces. We further release our reconstructed mouse brain model to the public. 
\item \emph{Virtual slicing of 3D brain surface:} We develop a user-interface which takes the 3D virtual mouse brain model as input along with the slicing angle and slicing interval to produce a series of annotated atlas slice images corresponding to microscopic slices sliced in the same direction and slicing interval. 
\item \emph{Robust edge detection algorithm:} We propose a novel edge detection algorithm, which is a variant of the Canny edge detector, to detect strong edges in the microscopic as well as annotated slice images. These edges are then used as features for the inter-stack registration.
\item \emph{Damaged region detection:} We automatically detect damaged regions in individual slices using Constrained Delaunay triangulation (CDT) and remove the features corresponding to these damaged regions for robust registration. 
\item \emph{Non-linear registration:} We perform non-linear registration of the mouse brain microscopic slices with the annotated atlas slices using Laplace equations with Dirichlet boundary conditions and validate its accuracy both qualitatively and quantitatively. 
\item \emph{Testing:} We show the robustness of our algorithm through accurate alignment of over 200 mouse brain slices with and without artifacts from various parts of the brain, and on data sets acquired through two different imaging techniques.
\item \emph{Application:} As an application of our registration, we perform region based neuron counting in a single mouse brain dataset by segmenting the neurons from microscopic slice images and counting the number of neurons in all the 20 3D anatomical regions.
\end{itemize}

\vspace{-0.1 cm}
\section{Related Works}
\label{sec:relatedwork}
In this section we discuss the previous works that are most closely related to our proposed approach. 

\noindent
\textbf{3D Surface Reconstruction of Mouse Brain:} Building a surface from curve networks (multiple or nested curves) on parallel slices and guaranteeing a valid geometry (manifold with no self-intersections) and correct topology (same topology as the original object) is a non-trivial problem \cite{Ju:2005aa}. Although there have been several works in this direction, most of them have the user check the validity of geometry and topology of the structure and make appropriate changes. 

Ju et al. \cite{Ju:2005aa} reconstructed the mouse brain with 17 anatomical regions using 350 pre-aligned coronal histology slices having a small inter-slice distance of 25 $\mu$m. They proposed a projection based approach, where contours on parallel slices are projected onto a common plane, afterwhich a volume graph is constructed using the intersections of these projections before triangulating the surface. A similar strategy in \cite{Liu:2008aa} projects the contours onto the medial axis followed by mesh refinement to obtain a smooth reconstruction. This method can also handle curve networks on non-parallel contours and shows the reconstruction of the mouse brain with 10 anatomical regions from 14 non-parallel contours. A drawback with both these \emph{projection-based} approaches is that the reconstruction heavily depends on the configuration of the cross-sectional planes and the inter-slice distance:  if during projection, there is no intersection between the contours of the same region from two adjacent slices, that region would be disconnected in the final reconstruction. Hence, the topology of a region varies greatly by changing the input configuration of slices. Zou et al. \cite{Zou:2015aa} later introduced a method through which the user can control the topology of the final reconstruction by using an additional 3D intensity volume as input. Another common approach is to first create a volume - voxelize the contours on each plane and stack them together \cite{Weinstein:2000aa,Kuan:2015aa}. This is usually followed by iso-surface extraction. Although the surfaces generated are geometrically correct, since it depends on regular voxel grids, the surfaces generated only approximate (and not interpolate) the original contours.

Our problem of reconstruction of the mouse brain from ARA slices differs from the above approaches in two aspects. First, the ARA contains 132 coronal annotated slices which are not pre-aligned. Hence, before performing any reconstruction, all 132 slices need to be aligned such that the overall shape and size of individual regions are preserved. Second, as the inter-slice distance between ARA is large (100 $\mu$m), contours of same regions across neighbouring slices have large displacement making projection based approaches challenging.

\noindent
\textbf{Image Registration:}  Image registration essentially consists of placing two images or volume datasets, acquired using the same or different imaging modalities, into a common coordinate system such that all the relevant features are aligned. There is a huge body of work on registration of medical images and the readers are referred to \cite{Maintz:1998aa} for a detailed categorization. Here we briefly discuss one classification most relevant to our approach. Registration can broadly be classified as \textit{intra-stack} (registration among slices within a stack) or \textit{inter-stack} (registration among slices in between two stacks). Further, within each of them there exists both \textit{feature based} and \textit{intensity based} registration techniques.  

\textit{Intra-stack} registration of mouse brain slices using either features \cite{Hess:1998aa,Zhao:1993aa,Rangarajan:1997aa} or intensity \cite{Nikou:2003aa,Zhao:1993aa} typically assume little distortion between consecutive slices and hence either use a rigid, affine, or similarity transform to get a global smooth 3D alignment. Ju et al. \cite{Ju:2006aa} proposed an intensity based method called warp-filtering which models the 2D deformations by decomposing them into two 1D deformations, one for each horizontal and vertical directions. To handle slices with small artifacts like air bubbles and tissue folds, they average pixels corresponding to the artifacts from neighbouring slices. Although such pixel averaging might achieve a smooth 3D volume, it is inadequate for accurate intra-stack alignment. Kovacevic et al. \cite{Kovacevic:2005aa} used existing intensity based methods to align low resolution MRI images of nine mouse brains to create a variational atlas - an atlas which encodes the variation of different anatomical regions as deformation field.

\textit{Inter-stack} registration between atlas and microscope slice images using intensity based approaches require the intensity profiles of both the microscope and atlas slice images to be same. This either constrains the experimental setup to perform same staining as the atlas or compels the neuroanatomist to create an intermediate atlas (either an average image \cite{Kuan:2015aa}, blockface photographs \cite{Dauguet:2007aa}, manual synthetic intermediate atlas \cite{Vousden:2014aa} or MRI images \cite{Ma:2005aa}) to aid in registration. Ng et al. \cite{Ng:2005aa}  proposed a technique for In-Situ Hybridization (ISH) to Nissl registration without using any intermediate atlas but using a region based deformable registration through warping of both Nissl reference and annotated atlas slices using high resolution B-spline grid. On the other hand feature based approaches for \emph{inter-stack registration} require extraction of reliable features from microscope images followed by alignment. Ali et al. \cite{Ali:1998aa} aligned only the outermost contour features (obtained by manual thresholding) using inflection points and area invariant descriptors by assuming a global affine transformation. Carson et al. \cite{Carson:2010aa} proposed a sub-division mesh based atlas to semi-automatically align ISH slices to the atlas. During this procedure, they first manually construct the sub-division mesh for 11 saggital slices from Valverde atlas \cite{Valverde:1998aa}. This mesh is then later fitted onto saggital ISH slices using a combination of statistical shape model, anatomical landmarks, and region boundaries. The corresponding landmarks used during this fitting process were detected a priori via a classification method which computes features from regions arranged manually for individual landmarks. This model-to-image alignment was later improved by Le et al. \cite{Le:2012aa} using a Markov random field framework along with mutual information to compute the optimal location of control points for an accurate alignment.   
 
\emph{We perform feature based inter-stack non-linear registration between mouse brain microscopic slices and their corresponding annotated ARA slices.} We assume no prealignment of slices. Unlike earlier approaches, our method can handle slices that may have histological artifacts (tears, tissue loss and deformations), very common in conventionally produced slices. Using our edge detection algorithm, we align \emph{both} outer and inner contours for an accurate alignment. Aligning only the outer contours and propagating deformations inside does not achieve exact interior alignment. Since our method does not use intensity for registration, we do not need to create any intermediate atlas and our proposed method can align microscopic slices from different imaging modalities and staining procedures to ARA. Furthermore, since we directly use the annotated atlas slices, we only need to perform warping once, unlike previous intensity based techniques which first compute the deformation using Nissl atlas slices and then perform warping of the annotated atlas slices. 

Although serial two-photo tomography (STPT) produces artifact-free, well-aligned, high-resolution 3D datasets that makes the registration process much easier \cite{Lein:2007aa,Ragan:2012aa,Kuan:2015aa}, conventional techniques to process the brain, although may produce deformed and damaged brain slices, continues to be the protocol of choice in many laboratories around the world because of the flexibility it provides for post slicing analysis including staining. We present methods to automatically handle even such damaged slice images during our registration. 

\noindent
\textbf{Damage Region Detection:} Most previous methods which try to automatically detect slices with artifacts look for unexpected differences between a specified slice and its neighbouring slices \cite{Kindle:2011aa, Qiu:2009aa,Ju:2006aa}. In other words, artifacts in an isolated slice cannot be detected or corrected. Further, such a method also requires slices to be close enough and the adjacent slice to be devoid of any artifacts, such that the difference between slices will imply the artifact. This limitation sometimes restricts the neuroanatomists who may want slices only from specific regions of the brain or want to slice the brain at larger intervals. There have also been efforts to not only identify but correct these artifacts. Kindle et al. \cite{Kindle:2011aa} proposed a semi-automatic method where they manually identify small tissue tears and fill them by warping neighboring regions around the tear. This approach only works well when the tear is small, horizontal, and smooth. Further, one needs to be careful about obtaining undesirable warping effects while fixing these tears, especially when they are as severe as the ones shown in Figure \ref{fig:registrationMI}. 

While the above techniques aim to detect and correct slices which have artifacts, many researchers try to overcome them. They use methods such as cryosectioning of the frozen mouse brain \cite{Crecelius:2005aa,Lein:2007aa,Carson:2010aa}, where they embed the brain in gel like compounds making it much easier to slice tissues into thin sections without tears or significant deformations. Another method quite popular is the introduction of quality control checks \cite{Lein:2007aa, Yushkevich:2006aa}, where highly damaged slices are manually removed from the registration pipeline. A major problem with this approach is that if enough of such slices are removed, there may not be sufficient information left to register and reconstruct the 3D brain model. Further, to aid in registration of such highly damaged slices, manual landmarks are often placed \cite{Crecelius:2005aa} or even manual initial registration is performed \cite{Vousden:2014aa, sawiak2009aa}. All the above measures which mitigate the 2D slice-specific artifacts and help its registration, in addition to being time consuming, are expensive and require a lot of experimental planning. Although, slicing thicker sections may be a plausible solution to avoid tissue tears \cite{Berlanga:2011aa}, it constrains the subsequent staining and imaging procedures. One needs to ensure that the slicing thickness is in accordance with the penetration depth of the stain and depth of focus of the light microscope used. 

\begin{figure}[!t]
\centering
\includegraphics[width=0.49\textwidth,height=6cm]{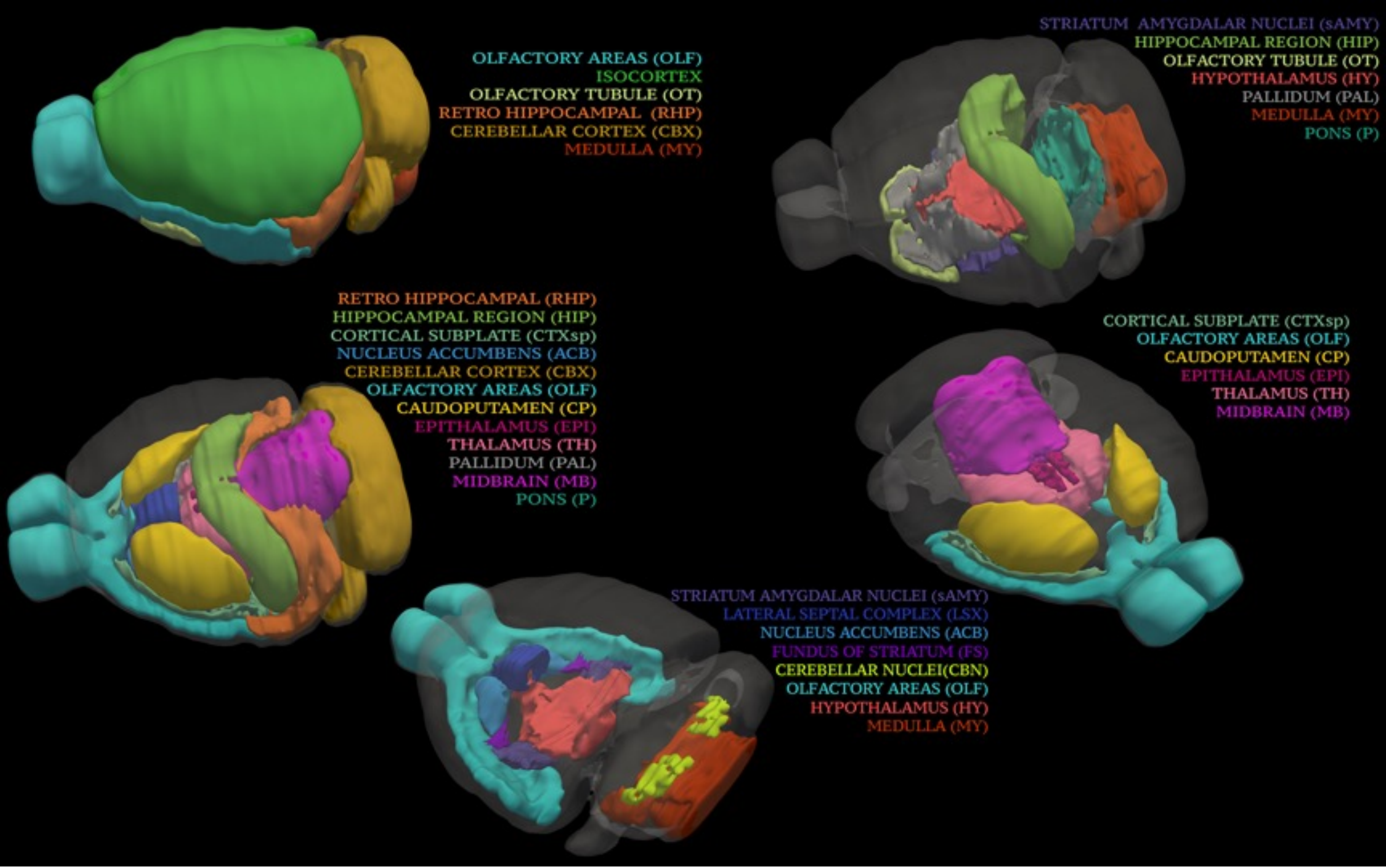}
\caption{\textbf{Result of our surface reconstruction:} Surface reconstruction of 20 major anatomical regions of the mouse brain where each 3D region is a smooth manifold mesh with correct topology and no intersections within or between regions. (please zoom in for details)}
\label{fig:mouseBrainModel}
\end{figure}

Our automatic method to detect damaged regions (histological artifacts) only requires the slice in question as input and does not use any information from the neighbouring slices. Furthermore, since we perform a feature based registration, we simply remove the features present in these damaged regions and hence avoid the need to correct these artifacts which might introduce undesirable warping effects.

\vspace{-0.1 cm}
\section{The Proposed Algorithm}
\label{method}

Our goal is to register mouse brain microscopic slices, sliced in any arbitrary direction and interval, to a standardized annotated 3D mouse brain model. Hence, we first reconstruct the 3D model using annotated atlas slices from ARA. This virtual 3D mouse brain model can then be sliced at the same orientation and interval as the microscopic slices, corresponding slices be registered with each other, and thus an annotated 3D common reference space is constructed where the annotations and 3D reconstruction can be easily transferred to the mouse brain as shown in Figure  \ref{fig:pipeline}.

\subsection{Surface Reconstruction of Mouse Brain Atlas}
We perform surface reconstruction of the mouse brain from 132 parallel coronal annotated atlas slices of ARA. Although each annotated slice is delineated into numerous regions, we only use and reconstruct 20 major anatomical regions in the mouse brain\footnote{See Table \ref{tab:neuronProfiling} for the full list}. These regions were chosen based on the hierarchical structure of the mouse brain tree provided by the Allen Brain Institute. Each of the 20 regions were reconstructed individually, smoothened, and intersections were semi-automatically resolved. Hence, the input to our surface reconstruction algorithm is 132 parallel coronal annotated atlas slices and the output is a 3D reconstructed mesh of 20 anatomical regions. We now describe the various challenges during the surface reconstruction procedure and solutions to address them. 

\subsubsection{Challenges in Surface Reconstruction}
\textbf{Relative shape scale preserving alignment:} The annotated slices from ARA maps are not aligned. Using such unaligned slices will introduce ripples in the 3D reconstructed brain model. Hence, our first step is to align all 132 coronal slices together while preserving the relative shape and size of individual regions between and within slices. Aligning the centers of the 
\begin{wrapfigure}{r}{0.2\textwidth}
\vspace{-2mm}
\centering
\includegraphics[width=0.2\textwidth,height=4cm]{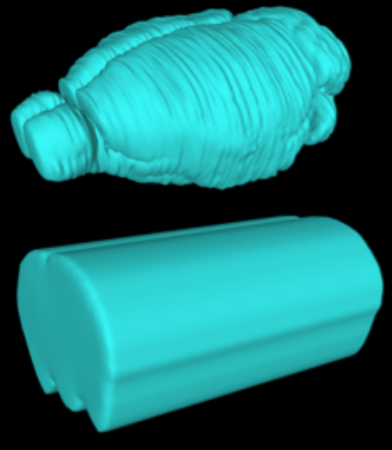}
\vspace{-6mm}
\end{wrapfigure}
annotated slices preserves the relative shape and size, but does not guarantee smooth reconstruction as shown from the ripples in the top inset figure. Aligning the contours (from both internal and external regions) guarantees smoothness of the 3D reconstructed shape, but does not preserve relative shape and size between slices and regions. The shape of the brain is lost as shown in the bottom inset figure. 

Since we want to preserve the relative shape and scale of all regions during alignment, we first align the centers of all the slices and then remove all those slices which are not aligned with the help of the subject expert. Using the remaining aligned slices, we then reconstruct the mouse brain surface via BPA (ball-pivoting algorithm) \cite{Bernardini:1999aa} as it connects neighbouring contours without introducing new vertices. Once the reconstruction is performed, we insert those unaligned slices which were earlier removed and compute the intersection of their plane with the surface of the reconstructed mouse brain. 
\begin{wrapfigure}{r}{0.2\textwidth}
\vspace{-2mm}
\centering
\includegraphics[width=0.2\textwidth,height=4cm]{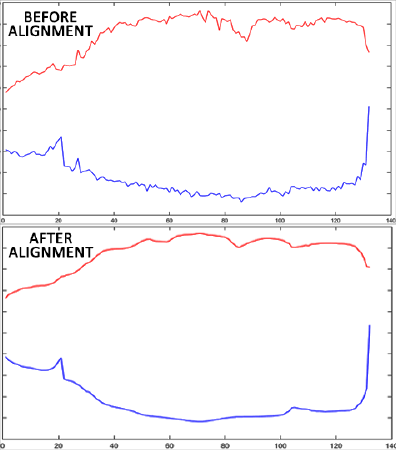}
\vspace{-6mm}
\end{wrapfigure}
Using the intersection values, we compute the required scale which is necessary to align each unaligned slice. We \emph{uniformly-scale} the unaligned slices such that the profile of all 132 slices is smooth when stacked on top of each other. The inset figure shows the profile of the top contour (red) and bottom contour (blue) of the mouse brain before (top) and after alignment (bottom). 

\noindent
\textbf{Surface Reconstruction \& Resolving Self-Intersection: } Once all the slices are aligned, we perform surface reconstruction to generate a 3D virtual mouse brain model. Annotated slices from ARA have large inter-slice distance making the use of previous projection based reconstruction approaches \cite{Ju:2005aa,Liu:2008aa} difficult as they would not guarantee topological correctness for all the regions. 
\begin{wrapfigure}{r}{0.2\textwidth}
\vspace{-2mm}
\centering
\includegraphics[width=0.2\textwidth,height=4cm]{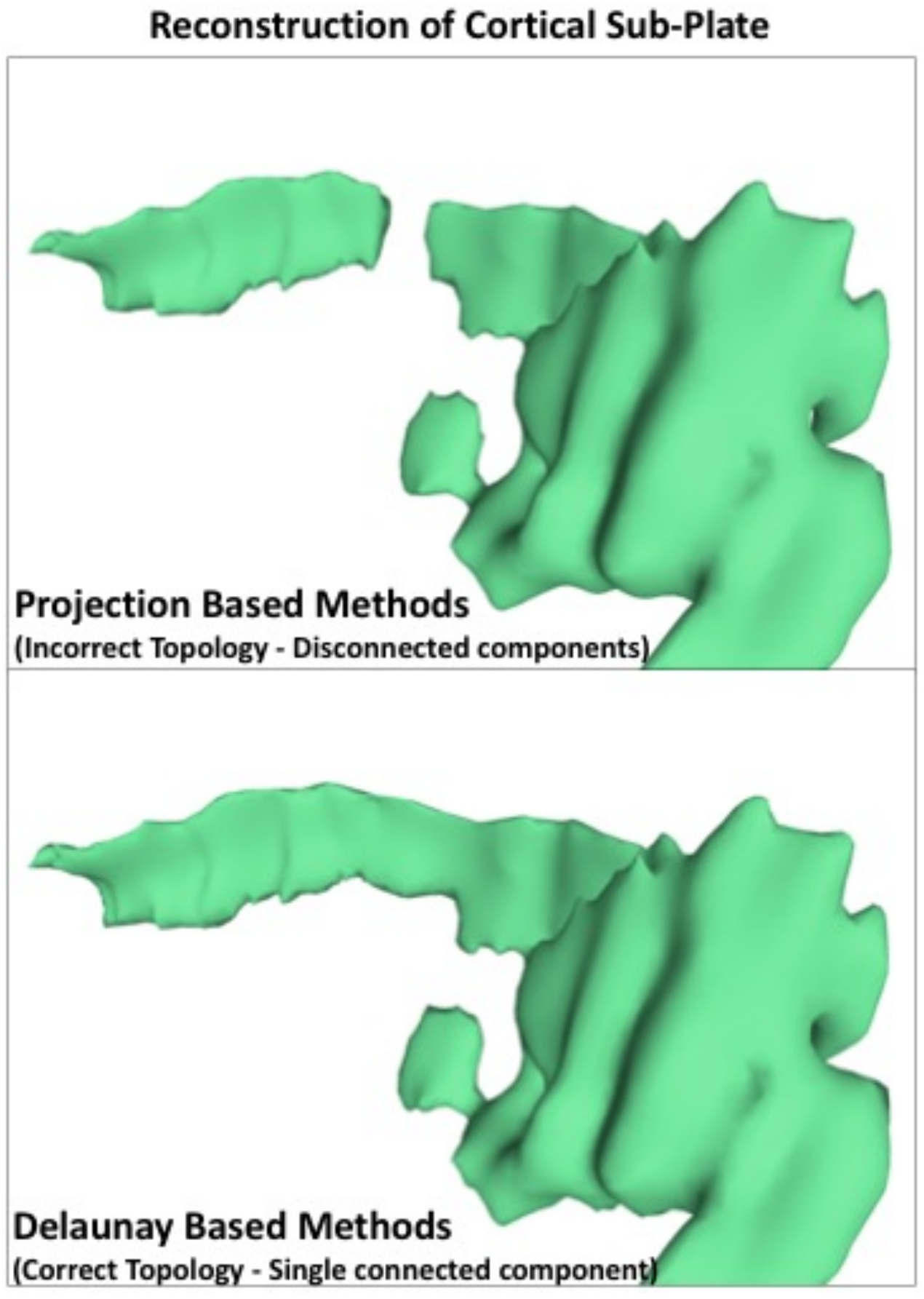}
\vspace{-6mm}
\end{wrapfigure}
A single anatomical region could be disconnected into multiple regions as there may not be any intersection between its contours from adjacent slices (top inset figure). To avoid these topological errors, we first automatically separate all anatomical regions in each annotated slice image based on their RGB color. We then reconstruct each region separately using Delanay triangulation \cite{Boissonnat:2007aa}, thereby forcing regions with contours that have large displacements to form a single connected component (bottom inset figure). We later merge all the individual reconstructed regions to obtain the final mesh. 

Although the final reconstructed mesh has correct topology, it may not have a valid geometry. As we reconstruct regions individually without any constraints from neighbouring regions, there may be intersection between two adjacent 3D reconstructed regions. Its well known in the literature that resolving such intersections (detection and correction) with minumum change to the intersecting regions is extremely challenging to automate as it depends a lot on the topology of the intersecting parts. Hence, we adopt a semi-automatic approach where we compute the intersecting triangles between the regions and remove those intersecting triangles along with any disconnected components which are small in size. The gaps thus created are then filled using an approach similar to \cite{Attene:2010aa}. This is followed by a surface fairing algorithm \cite{Taubin:1995aa}, which does not shrink the mesh. This results in the 3D model of the mouse brain where the 3D structure of each region is reconstructed, annotated, smooth and free from intersections. The reconstructed mesh for each region is a manifold for the ease of applying further geometry processing algorithms. Such manifolds are achieved, if required, by duplicating vertices and triangles of the shared boundary between two adjacent regions. Although the reconstruction process is labor intensive, this is a one time operation and we have released the final 3D reconstructed brain model for public use. 

\subsection{Registration of Annotated and Microscopic Slices}
Before we perform registration of the microscopic slices, we first need to compute their corresponding annotated atlas slices. We achieve this by slicing the reconstructed mouse brain model using the same slicing direction and slicing interval used to slice the actual mouse brain. This is important because the ARA is a 2D atlas with slices at very specific intervals and slicing angles, allowing for only an approximated (closest possible match) atlas slice to a given microscopic slice.  

Hence, given a stack of microscopic slice images, using our interface (please see supplementary document) we first slice the 3D virtual mouse brain model at the same slicing angle and slicing interval to generate annotated atlas slice images (AI) corresponding to each individual microscopic slice image (MI). Given these matching slices, the rest of the paper explains the procedure to register one MI to one AI. 

\subsubsection{Feature Extraction}
\label{ssec:featureExtraction}

The first step in our registration pipeline involves extracting features. For this we extract the dominant edges or contours from both AI and MI and also compute the orientated bounding boxes (OBB) \cite{Gottschalk:1996aa} around these dominant edges. Both these features are later used for accurate alignment. 

\begin{figure}[!t]
\centering
\includegraphics[width=0.49\textwidth,height=6cm]{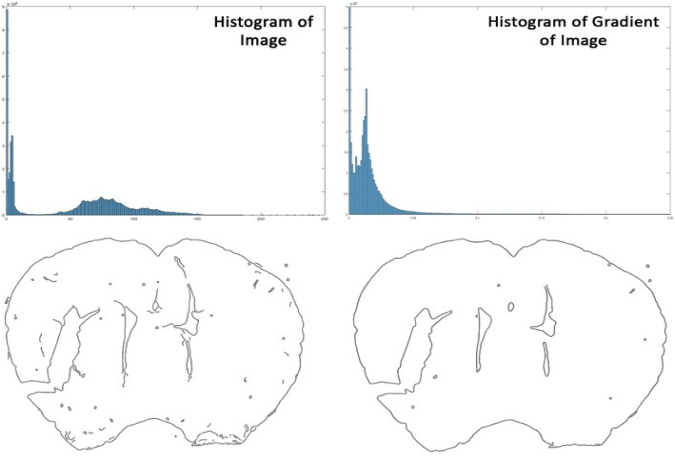}
\caption{ \textbf{Edge detection on a microscope image:} The histogram of the microscope images do not display a clean bimodal distrbution causing Ostu's method to generate noisy edge images (left). In comparison, our method uses the histogram of the gradient magnitude to generate cleaner edge images (right). (please zoom in for details)}
\label{fig:edgeDetect}
\end{figure}

We first create an \textit{atlas-edge image} (AEI) by extracting the edges from AI. In order to compute the \textit{microscopic-edge image} (MEI) from MI, whose edges correspond to edges in AEI, we propose a novel \emph{dominant edge detection} (DED) algorithm that is a variant of the Canny edge detector. The DED algorithm automatically computes the threshold for hysteresis as described in Algorithm \ref{algo:EdgeDetect}. Ostu's method used in Canny edge detector, requires the input image to have a bimodal distribution as it computes the threshold for hysteresis by maximizing the inter-class variance \cite{otsu1975aa}. The MI generally do not display a clean biomodal distribution (Fig. \ref{fig:edgeDetect}) and hence the threshold value computed by Ostu's method is not stable - a lot of spurious edges are added with small errors in the threshold value. In comparison, our DED algorithm uses the idea of \emph{persistence of edges} from the histogram of the gradient magnitude. We compute the threshold by finding a stable region or interval on the histogram of gradient magnitude. The intuition behind our algorithm is to compute a threshold value which not only suppresses weak edges, but also introduces fewer spurious edges when small errors are present. Our threshold computation algorithm performs better than the standard Otsu's method as it removes small weak edges (Fig. \ref{fig:edgeDetect}) which potentially could lead to wrong correspondences during putative matching. 

\begin{algorithm}[!b]
\caption{Edge threshold ($K$) computation for input MI}
\label{algo:EdgeDetect}
\begin{algorithmic}[1]
\State Smooth MI with isometric median filter of size $w_m$ and Gaussian filter with standard deviation $\sigma_g$ and size $w_g$. 
\State Compute the histogram of the gradient magnitude.
\State The number of bins $b$ in the histogram is computed using Scott's rule \cite{Scott:1979aa}, 
$b=3.49\sigma_s N^{-\frac{1}{3}}$ where $\sigma_s$=standard deviation of the $N$ gradient magnitude values.
\State Compute first $k$ bins such that the difference of number of points in adjacent bins of the $k$ bins lie within a fixed threshold $s$.
\State \Return $K$= Mean of gradient magnitude value corresponding to $k$ bins. 
\end{algorithmic}
\end{algorithm}

\begin{figure*}[!t]
\centering
\includegraphics[width=0.95\textwidth,height=6cm]{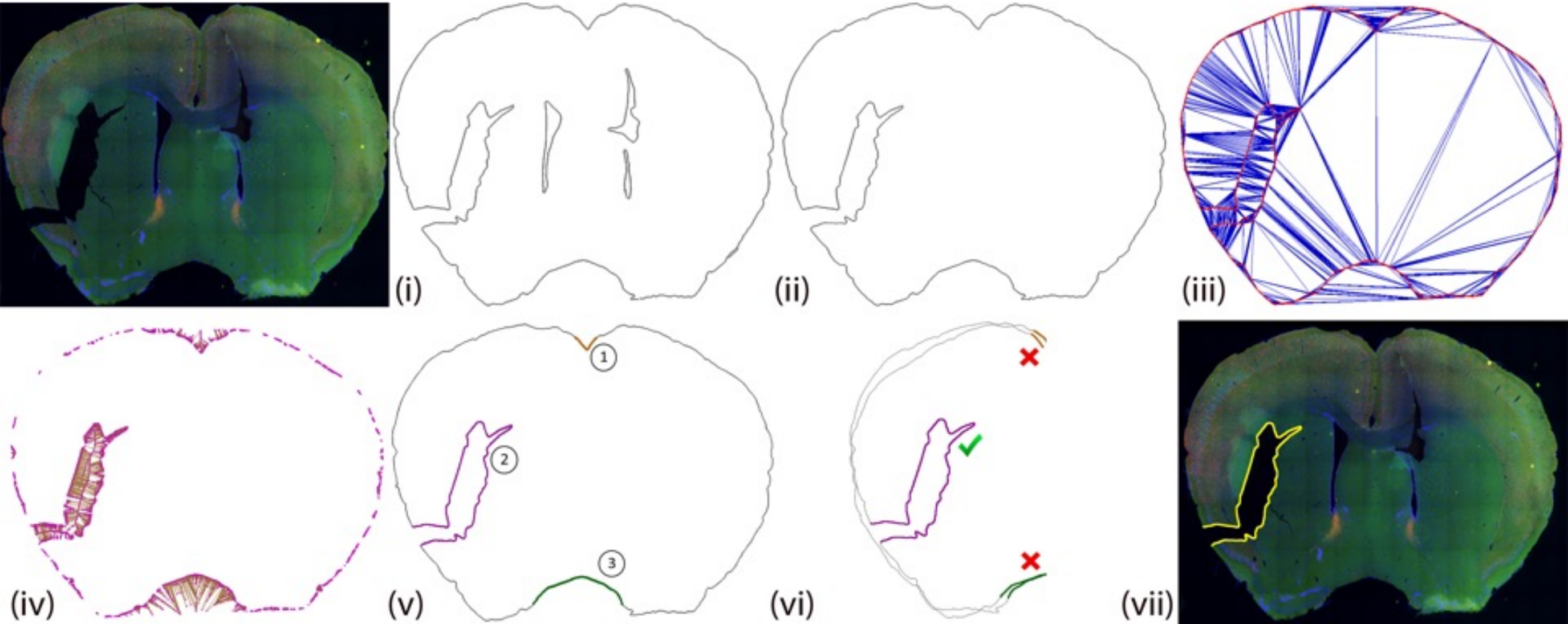}
\caption{\textbf{Overview of our damaged region detection algorithm:} (i). Dominant edges (MEI) extracted from the mouse brain microscopic slice image on the left. (ii). Outermost contour of MEI, which serves as the input to our algorithm. (iii). Constrained Delaunay Triangulation of vertices $V$ \& edges $E$ using the outermost contour of MEI. (iv). Exterior Voronoi vertices (magenta) and edges (brown). (v). Three candidate damage regions whose medial axis (Voronoi edge sequence) length was above $\alpha$. (vi) Candidate damaged regions vertically reflected about OBB and checked for symmetry. Points corresponding to only the 2nd candidate area were classified as damage region points as they were not vertically symmetric. Whereas points corresponding to other two regions represent features of the brain. (please zoom in for details)}
\label{fig:damageRegion}
\end{figure*}

Algorithm \ref{algo:EdgeDetect} computes threshold $K$ from MI. After removing noise and computing the histogram of gradient magnitude, we find the first consecutive $k$ bins, where the number of points remain stable within a fixed threshold $s$. Intuitively, for strong edges, the number of points in the nearby bins will not fluctuate much as compared to weak edges providing a large stable range. To compute the threshold ($K$) we then take the mean of the gradient magnitude values of these $k$ bins. Lowering these values, lowers the threshold thereby introducing weak or spurious edges. Hence, these parameters control how clean the MEI is, which in turn affects the correspondences and alignment.

Using both the microscopic and atlas edge image, we now compute their OBB.  However, computation of OBB using principal component analysis (PCA) will fail when used on highly damaged MI slices (Fig. \ref{fig:damageRegion}) because the spurious edge points produced in damaged areas of the tissue images bias the PCA. This is one of the reasons why other edge based registration algorithms could not handle serious tissue damage artifacts. We address this challenge in two steps. First, we approximate the bounding box using convex hull which eliminates the effect of internal tissue damage artifacts. Then we uniformly sample the convex hull to eliminate the sample bias effect in PCA computation. Computing OBB from AEI is straightforward as it does not contain any histological artifacts. 

\subsubsection{Detection of Damaged Regions}
\label{ssec:damage}
Before computing corresponding points between MEI and AEI, and using those to further align the two images, it is important to first accurately identify and remove points in the damaged regions. The presence of edges due to the damage regions misleads and corrupts the correspondences (Fig. \ref{fig:corruptCorres}), resulting in bad registration.

Our algorithm to detect damage regions in mouse brain slices is motivated by two key observations. First, contours of most damaged regions have long exterior medial axis creating deep concavities into the tissue (Fig. \ref{fig:damageRegion}). It is quite rare that the tear happens in the interior of the tissue directly without affecting the boundary of the tissue. Second, the damage region exhibits vertical asymmetry between the left and the right regions of the mouse brain. It is also very rare that the same type and shape of tear or missing region happens on both sides of the brain tissue slice.

\begin{algorithm}[!t]
\caption{Detection of points $P_D$ in the damaged regions in input MEI.\\  
\textbf{INPUT:} Vertices $V$ $\&$ Edges $E$ from outermost contour of MEI and $\alpha$.\\ 
\textbf{OUTPUT:} Points $P_D$ in the damaged regions of MEI}
\label{algo:DamageArea}
\begin{algorithmic}[1]
\State Construct a CDT using $E$ $\&$ $V$. 
\State Remove all the triangles inside the polygon formed by $E$. Also remove all the sliver triangles whose circumcenter does not lie inside their triangle.
\State Using the remaining $E$ $\&$ $V$, construct the dual Voronoi diagram. 
\State Compute all the Voronoi edge sequences $\geq$ $\alpha$ and let the triangle vertices $V$ corresponding to the remaining Voronoi vertices be $V^\prime$. 
\State Check for vertical symmetry $\forall$ $v \in$ $V^\prime$ and remove symmetric vertices from $V^\prime$.
\State \Return $P_D$ $\Leftarrow$ $V^\prime$ which are asymmetric. 
\end{algorithmic}
\end{algorithm}

Algorithm {\ref{algo:DamageArea} computes points $P_D$ in the damaged regions in MEI. Using the vertices $V$ and edges $E$ of the outermost contour of MEI, we first construct a Constrained Delaunay Triangulation (CDT). All edges of $E$ are a part of this triangulation as shown in red in Figure \ref{fig:damageRegion}(iii). By computing the winding number of a point inside the triangle, we then remove all the triangles lying inside the contour (winding number = 0) and retain only the exterior Delanuay triangles \cite{Jacobson:2013aa}. In order to obtain reliable Voronoi vertices and edges that can be used in computations downstream, we further clean the remaining exterior triangles by removing all ``skinny" triangles -- any triangle whose circumcenter does not lie within the triangle. From the remaining vertices $V$ and edges $E$ of the Delaunay triangles, we represent the exterior medial axis as the sequence of Voronoi edges that do not intersect the edge set $E$ \cite{Amenta:1998aa}. Since this would create many small medial axes as shown in Figure \ref{fig:damageRegion}(iv), we threshold them (remove edge sequence $< \alpha$) and retain only those medial axes corresponding to deep concavities. The vertices of the Delaunay triangles corresponding to the retained medial axis Voronoi vertices ($V^\prime$) serve as candidates for the damaged regions as shown in Fig \ref{fig:damageRegion}(v). There may be important features of the brain that may also have long medial axis, but these features are also symmetric on both sides of the brain. Hence, as the final step of our algorithm, we check whether the damage region candidate edge points are symmetric between the left and right half of the mouse brain: the points in the candidate damaged regions are reflected along the vertical axis of the OBB (computed in Section \ref{ssec:featureExtraction}) and for every reflected point, a small 3x3 neighbourhood region is checked for vertices in the original data set with similar normal vectors. If no such points are found, then it is declared that there is no symmetry, the points in the identified region are classified as damaged area points and removed from MEI (Fig. \ref{fig:damageRegion}(vi)). 

\subsubsection{Non-Linear Registration}
\label{ssec:nonlinear}

To accurately align the microscopic image with its corresponding annotated atlas image, we first perform global affine alignment using ICP followed by local non-linear alignment by solving Laplace's equation with Dirichlet boundary conditions. 

\noindent 
\textbf{Global Alignment using ICP:} Before performing global alignment, we first resolve the rotation component between the two images as accurately as possible. 
This is achieved by computing the relative translation, scaling, and rotation required to align the OBB of both AEI and MEI. 

As both MEI and AEI are now coarsely aligned, we assume that the rotation component is resolved and only the translation and scaling components needs correction. Hence, for corresponding points on the edge curves of AEI and MEI, we can assume that the normal vectors would be almost the same. We compute the normal vectors of the points in AEI and the remaining points (after removing the damaged regions) in MEI using moving least squares \cite{Levin:1998aa} as it smoothly interpolates the normal vectors, diminishing the effect of noise, sharp features and topological foldings. Using these normal vectors, we then search for corresponding points between AEI and MEI within a small angle threshold. We may find multiple points $(x_a,y_a)$ in the AEI in a small neighborhood $\Omega$ corresponding to a single point $(x,y)$ in MEI. We assign weighted average of these multiple matches in AEI based on its Euclidean distance ($d_{ma}$) from the point in MEI, as the target position $(x^\prime,y^\prime)$ to which $(x,y)$ should finally be moved:

\begin{equation}
\label{weightedavg}
(x^\prime,y^\prime) = \frac{\sum_{i=1}^n w_i(x_{a_i},y_{a_i})}{\sum_{i=1}^n w_i}\hspace{8mm} \forall (x_a,y_a) \in \Omega 
\end{equation}
where
\begin{gather*}
w = 
\begin{cases}
100, & \text{if } d_{ma} = 0 \\
1 / d_{ma},  & \text{otherwise}
\end{cases}
\end{gather*}

\noindent
To exclude incorrect matches, we check if points in the neighborhood of a point in MEI are matched to the points in the same neighborhood in AEI. Using these robust correspondences, the affine transformation matrix $T$ is computed using linear least square formulation that would take the points in MEI $(x,y)$ as close as possible to their corresponding matching points in AEI $(x^\prime,y^\prime)$ as described below. 

Let $Z$ be $c$ x 3 matrix where $c$ is the number of correspondences in MEI whose target coordinates in the AEI are $x^\prime$ and $y^\prime$. Then $T$, the global affine transformation matrix can be computed by solving the linear equations, $ZT_1 = b_x$ and $ZT_2 =b_y$ where, 

\begin{equation*}
\label{affineequ}
 Z= \left[
\begin{array}{ccc}
x_1 & y_1 & 1\\
x_2 & y_2 & 1\\
.   & .   & .\\
.   & .   & .\\
x_c & y_c & 1
\end{array}
\right]  \hspace{5mm}
b_x= \left[
\begin{array}{c}
x_1^\prime\\
x_2^\prime\\
.\\
.\\
x_c^\prime
\end{array}
\right] \hspace{5mm} 
b_y= \left[
\begin{array}{c}
y_1^\prime\\
y_2^\prime\\
.\\
.\\
y_c^\prime
\end{array}
\right] 
\end{equation*}

\noindent
and $T_1$ and $T_2$ are the rows of $T$.

The global transformation thus computed may have non-uniform scaling, shear, and possibly a minor rotation adjustment component too. So, this transformation would change the normal vector of the points in MEI, which would lead to a slightly different set of matching points from AEI in the subsequent iteration, and potentially a different transformation matrix that would take MEI points further close to their new matches. Since we also use the distance of AEI points from the MEI (in the aligned images) for pruning the matching set of points, this \emph{iterative closest point} optimization will converge. We progressively use tighter normal vector angle deviation and smaller distance thresholds in subsequent iterations for quicker convergence.

\begin{figure}[!t]
\centering
\includegraphics[scale=0.25]{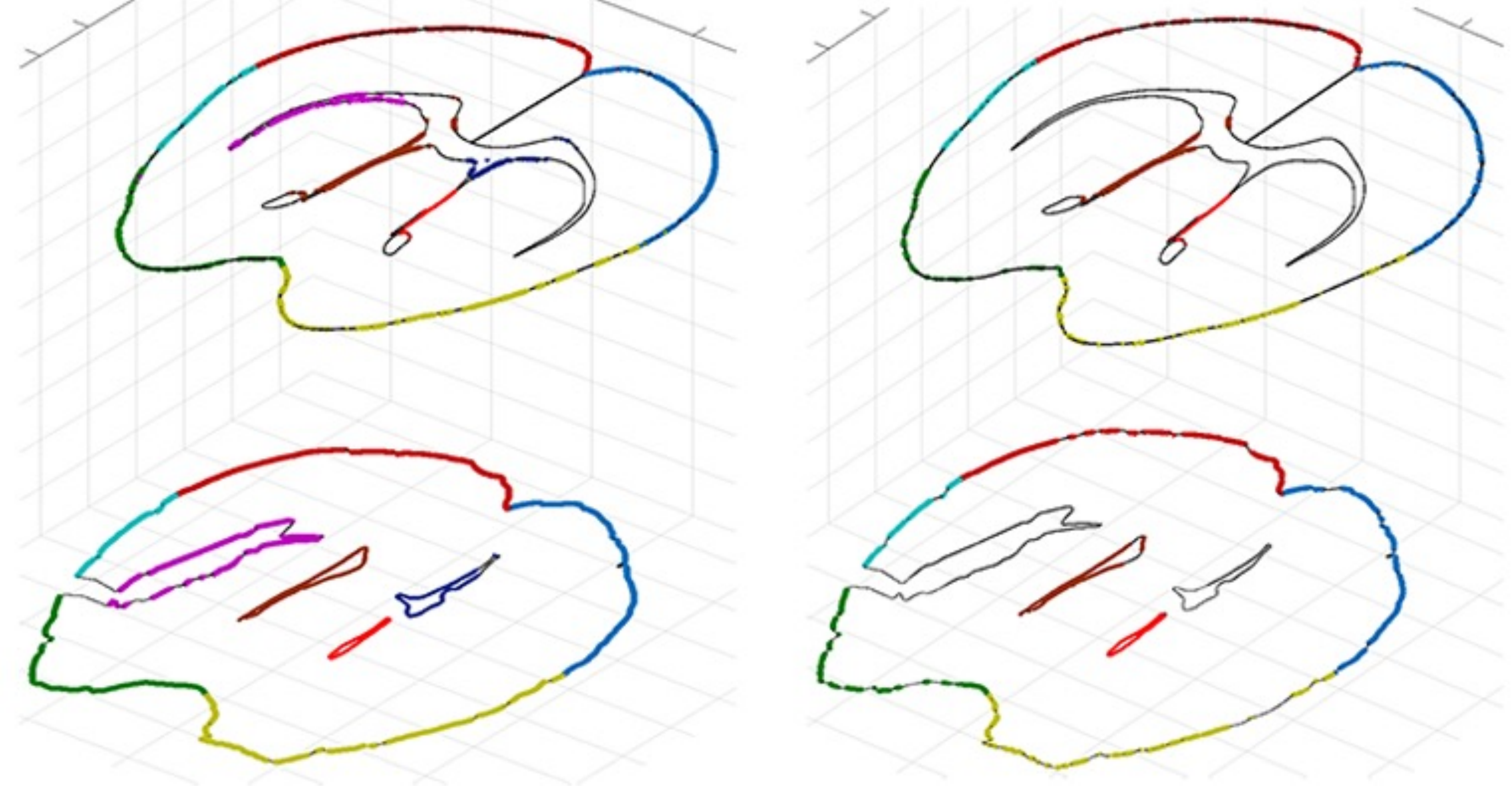}
\caption{\textbf{Correspondences used during MI-AI alignment:} Dense correspondences before (left) and after (right) outlier removal (damage region + incorrect correspondences) between MEI (bottom) and its corresponding AEI (top). Correspondences are shown using similarly colored curve segments. Note the damaged regions in MEI are incorrectly matched to features in AEI (left). These incorrect matches are removed using our damage detection algorithm (right).}
\label{fig:corruptCorres}
\end{figure}

\noindent
\textbf{Local Alignment using Laplacian:} After global affine transformation, we compute the final list of corresponding points between MEI and AEI that are spatially $\left(\frac{1}{40} \right)$ of image-height pixels apart, and deviate no more than 1 degree in their normal vectors. Given such tight correspondences, the next step is to register these points with each other using non-linear alignment technique. 

Let points in MEI, $P_M$, whose corresponding points in AEI, $P_A$, be given. This image warping function, $\phi(x,y)$, posed as the solution to Laplace's equation, should take each point in $P_M$ to its corresponding point in $P_A$. For points in $P_M$, this function is given as Dirichlet boundary condition:

\begin{figure*}[!ht]
\centering
\includegraphics[width=0.99\textwidth,height=9cm]{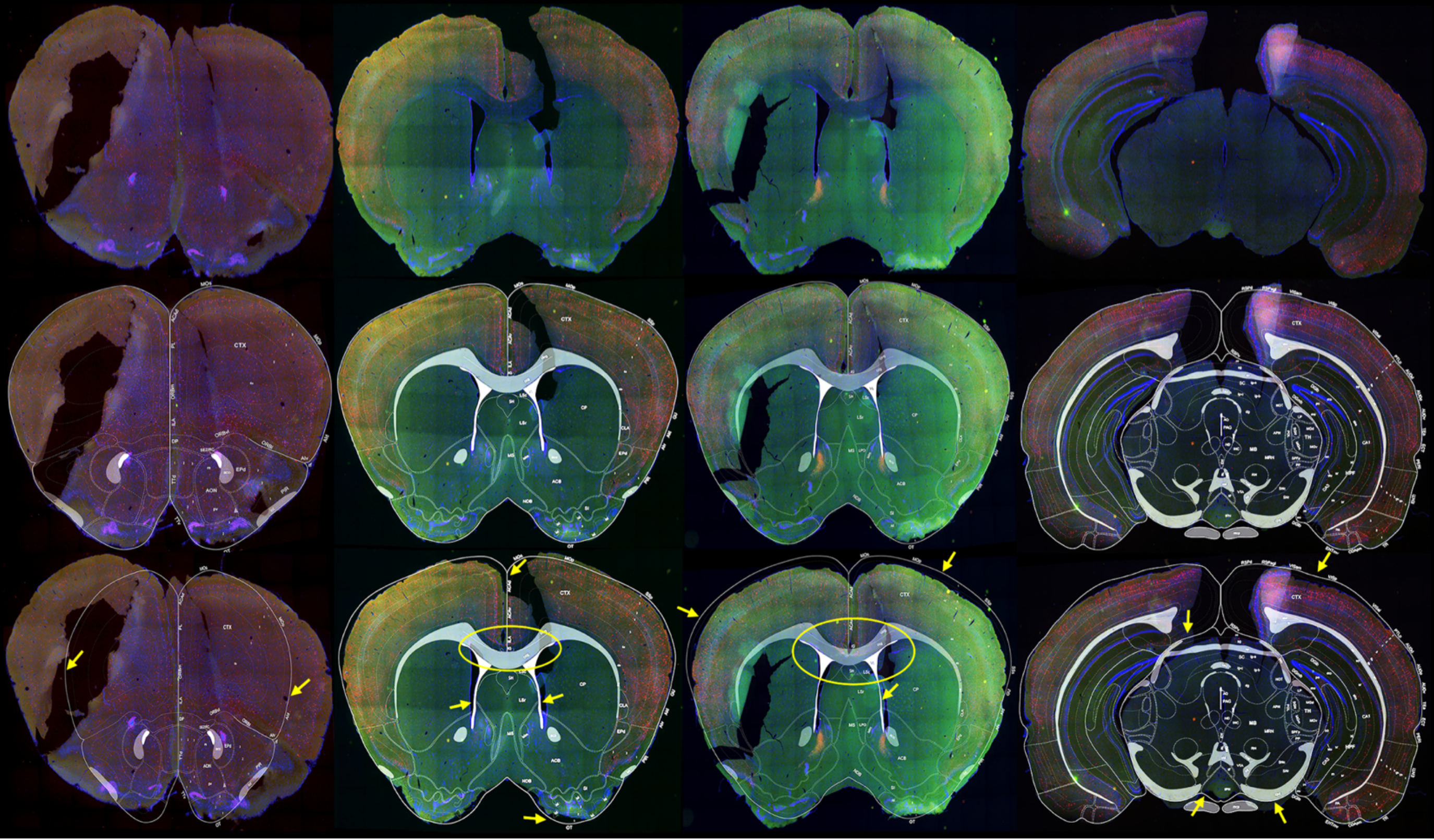}
\caption{\textbf{Comparison of registration results on damaged slices:} First row shows damaged coronal mouse brain slices produced from conventional histological techniques. Second row shows results from our registration algorithm on the these slices with the corresponding atlas overlayed in white. Third row shows results from affine + non-linear B-spline registration using Elastix \cite{Klein:2010aa}. A few sample locations of incorrect registration in the third row are shown using yellow arrows and marked regions. See supplementary document for more examples. (please zoom in for details)}
\label{fig:registrationMI}
\end{figure*}

\begin{equation}
\phi(s,t)=(B_x(s,t),B_y(s,t)) \hspace{2mm}  where \hspace{2mm} (s,t) \in P_M
\end{equation}

\noindent
where $B$ is the displacement vector between the corresponding points. Other pixels are distorted as little as possible by this warping function: 

\begin{equation}
\phi(s,t)=0 \hspace{8mm} \forall (s,t) \notin P_M
\end{equation}

\noindent
The smoothness in warping is achieved by the following Laplace's equations:

\begin{equation}
\nabla\phi_x  =  \frac{d^2\phi_x}{dx^{2}} + \frac{d^2\phi_x}{dy^{2}} = 0 \hspace{4mm}
\nabla\phi_y  =  \frac{d^2\phi_y}{dx^{2}} + \frac{d^2\phi_y}{dy^{2}} = 0
\end{equation}

\noindent
By approximating the second derivative at nodal point $\left(x,y\right)$ (derived from Taylor series), the finite difference approximation of Laplace's equation for interior regions can be expressed as a homogeneous system of linear equations of form 

\begin{multline}
\label{equ:Coeff}
\phi\left(x,y\right) = \frac{1}{4} \bigl( \phi\left(x+1,y\right) + \phi\left(x-1,y\right) + \\
\phi\left(x,y+1\right) + \phi\left(x,y-1\right) \bigr) = 0
\end{multline}

Combining the above equations and representing it in matrix notation gives $A\phi_x = C_x$ and $A\phi_y = C_y$ where $A$ is a $m\times m$ matrix and $m$ is the number of pixels in MI. The row vectors of $A$ takes the coefficients of terms in Equation \ref{equ:Coeff} except for the rows corresponding to MEI points in which case, it represents the Dirichlet boundary condition. The vector $C_x$ and $C_y$ are zero everywhere except for the rows corresponding to MEI points in which case it is $B_x$ and $B_y$ respectively. Note that $A$ is a sparse matrix allowing for efficient computation of the solution of $\phi$ that minimizes the residual, $ \parallel C- A\phi \parallel$.

The aforementioned algorithm is used to align each of the slices in the given microscopic stack to their corresponding annotated atlas slices. Using these registered microscopic slices and the reconstructed mouse brain, we create an annotated 3D common reference space which allows us to transfer 3D structure and annotations from the mouse brain model to the microscopic stack.

\vspace{-0.1 cm}
\section{Implementation \& Results}
\label{result}

\noindent
\textbf{Implementation:} The complete algorithm to align a single mouse brain microscopic slice which may have histological artifacts to its corresponding annotated atlas slice from ARA, takes about 1 minute with our unoptimized MATLAB implementation on an Intel Core i5 CPU with 8GB RAM. To compute MEI from MI, we first smoothened the MI using an isometric median filter with $w_m$=20 and Gaussian filter with $w_g$=12 and $\sigma_g$=2. After which, the threshold for hysteresis was computed using $s=12$, for which $k=5$. From the MEI, the damage region detection was performed with an $\alpha$ value of 20 to remove all the small medial axis's. The edges lying in the damaged regions were removed before performing global affine alignment using ICP with normal vectors as features. We progressively used tighter normal vector angle deviation thresholds $\left(10^{\circ}, 8^{\circ}, 6^{\circ}, 4^{\circ}\right)$ and smaller distance threshold $\left(\frac{1}{10},\frac{1}{20},\frac{1}{40},\frac{1}{80} \right)$ of image-height in each iteration for quicker convergence. 

\noindent
\textbf{Results:} We evaluate our registration pipeline on 200 coronal mouse brain microscopic slices (5000 x 8000 pixels) with a resolution of 0.6$\mu m$ per pixel. To test the robustness of our method, these images were taken from different datasets spanning different regions of the brain. Of these, 60 slices were from STPT \footnote{Publically available from the Allen Brain Atlas Project} (with no major artifacts) and the rest 140 produced from conventional processing techniques \cite{Sun:2014aa,Oh:2014aa} with many artifacts. From these 140 slices, 52 slices had histological artifacts (45 slices with single, and 7 slices with multiple artifacts) such as tissue tears and missing regions, which were produced either during serial sectioning of the mouse brain tissue or during manual mounting of the thin slices on the glass slides. 

\begin{table*}[!t]
\caption{Comparison of registration errors (in pixels) after affine $\&$ final non-linear (affine+elastic) transformations using intensity-based and our feature-based method.}
\centering
\begin{tabular}{|R| W | N N N | N N N|}
\hline
\multicolumn{2}{|c|}{} & \multicolumn{3}{c|}{\textbf{Clean Slices} (88 slices)} & \multicolumn{3}{|c|}{\textbf{Damaged Slices} (52 slices)} \\
\hline
\multicolumn{2}{|c|}{} & Average RMSE & Average MEE & Average MAE & Average RMSE & Average MEE & Average MAE\\
\hline
\smallskip
\multirow{2}{8em}{After Affine Transformation} & Intensity-Based & 11.37 & 8.89 & 24.88 & 19.56 & 12.13 & 40.71 \\
& Proposed & 12.78 & 10.90 & 23.42 & 13.22 & 10.37 & 24.81 \\ \hline
\multirow{2}{8em}{After Non-Linear Transformation} & Intensity-Based & 4.18 & 3.02 & 7.43 & 6.98 & 5.85 & 22.51 \\
& Proposed & 3.62 & 2.25 & 4.8 & 3.62 & 2.07 & 5.42  \\  \hline
\end{tabular}
\label{tab:registrationResults}
\end{table*}

\begin{figure}[!b]
\centering
\includegraphics[width=0.49\textwidth,height=5cm]{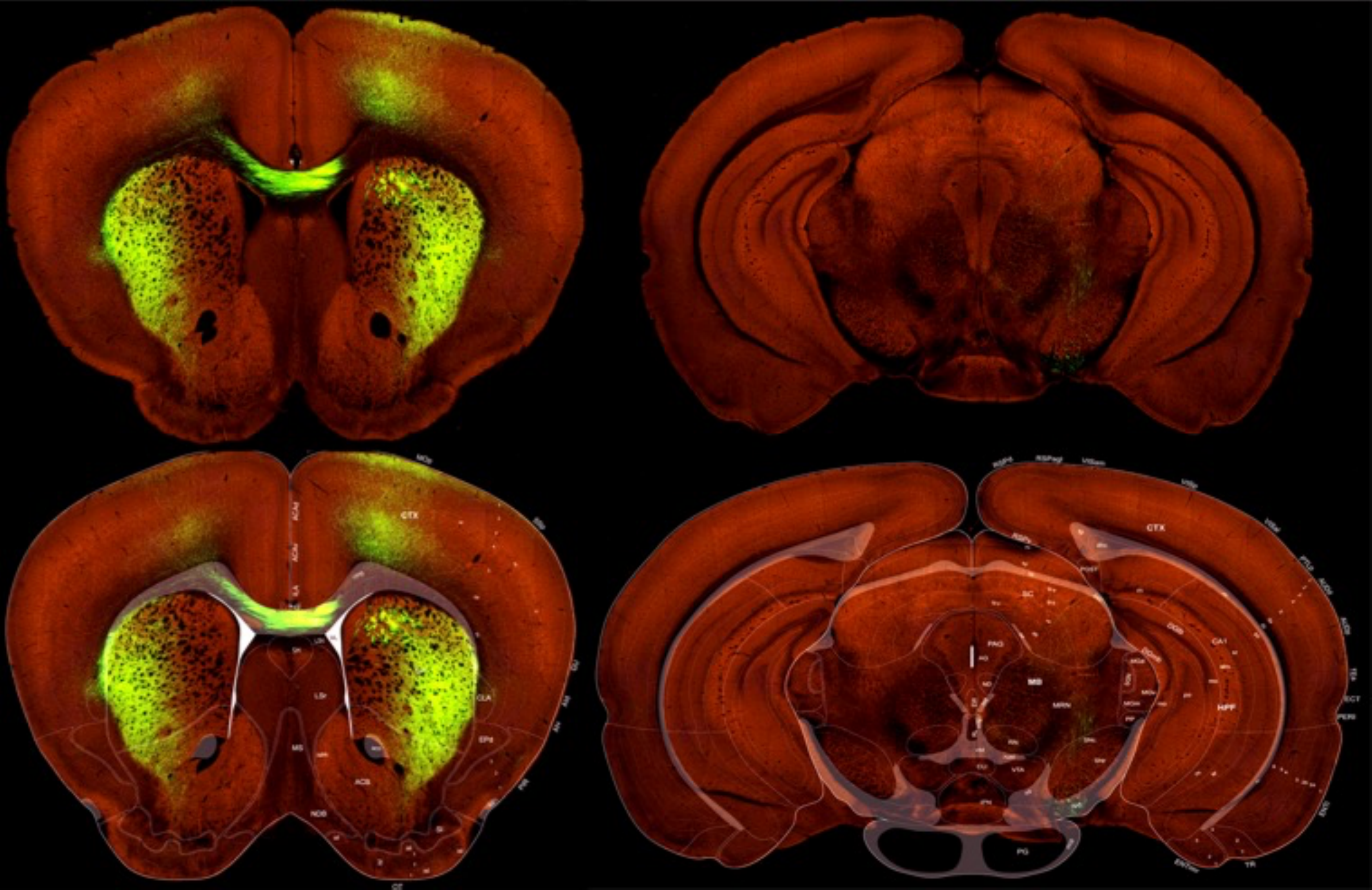}
\caption{\textbf{Registration results on clean slices:} The first row shows coronal slices from STPT while the second row shows results from our registration algorithm with the corresponding atlas overlayed in white. (please zoom in for details)}
\label{fig:registrationClean}
\end{figure}

Our registration results for all 200 slices (Fig. \ref{fig:registrationMI} \& \ref{fig:registrationClean}) were found qualitatively quite accurate by the subject experts. We also perform quantitative evaluation and comparison of our method with a similar end-to-end intensity based registration method that uses mutual information as its similarity metric. We chose mutual information because it is the most commonly used and popular metric for such \textit{inter-stack} registration problems (Microscope  to Atlas) \cite{Pluim:2003aa,Ng:2005aa, Ng:2007aa}. We implemented the above method in Elastix \cite{Klein:2010aa}, an ITK based modular framework, where we first started with precomputed optimized parameters (from Elastix) and later modified them for the best overall performance. We used Advanced Mattes mutual information to register the DAPI stained microscope images with the Nissl intensity images from ARA maps by performing an \emph{affine registration} followed by an \emph{elastic cubic B-spline} based transformation using a multi-resolution approach. An adaptive stochastic gradient descent optimizer with the final B-spline grid spacing of 18 pixel was used to ensure matching of local structures. We compared the root-mean-squared error (RMSE), the median error (MEE) and the maximal error (MAE) of 20 corresponding points which were manually picked and distributed uniformly in MI and AI pair. This comparison was done only for 140 slices from conventional processing techniques as there were no corresponding Nissl images for slices from STPT, hence they could not be registered with the Nissl intensity images from ARA maps. Although damage identification is done automatically, in order to collect statistics on results and for comparison with other methods, damaged slices were manually identified by subject experts, separated from clean slices, and separate comparisons were done on those slices.

There are two stages (affine $\&$ non-linear) to the pipeline and the two registration methods have different algorithms to realize each of these stages. While our method uses features to align both slices, registration performed by Elastix was done using Mutual Information - an intensity based similarity metric. The results in Table \ref{tab:registrationResults} are reported after each of the two stages. We found that during affine registration of damaged slices (52 slices), our proposed method gives lower registration errors (in terms of average RMSE $\&$ MAE) as compared to the intensity based method. For clean slices (88 slices) both performed equally well. After the non-linear registration, even in clean slices, we performed slightly better than intensity based method in all statistical measurements as we are using Laplace equations. Laplace equations, like thin plate spline (TPS) or B-spline also minimizes the total curvature. The addition of point correspondences as Dirichlet boundary conditions further constrains the interpolation of the displacement functions for an accurate MI to AI alignment.

Apart from registration, we also qualitatively validated our 3D reconstructed mouse brain model by subject experts. An accurate surface reconstruction of 20 major anatomical regions chosen from the Allen Brain Atlas project was performed from 132 parallel coronal annotated slices as shown in Figure \ref{fig:mouseBrainModel}. Our 3D mouse brain model is smooth with no intersections and each anatomical region is a manifold mesh.

\vspace{-0.1 cm}
\section{Application: Neuron Counting}
\label{sec:application}

As one of the application of our registration, we performed region based neuron counting in 51 microscopic slices from a single mouse brain dataset. After the alignment of all the MI to their corresponding AI slices, we transferred annotations from the AI onto the MI. We then performed segmentation of neurons in the registered MI slices followed by counting the number of neurons in all the 20 annotated regions. Figure \ref{fig:teaser} shows the 3D visualization of the neurons inside the reconstructed virtual mouse brain model.

The mouse brain used for neuron counting was labeled with Recombinant Adeno-Associated Virus (rAAV) expressing EGFP and counter stained with DAPI. This resulted in each microscopic slice to have 3 channels (8-bit) - EGFP projection signals were visualized in the green channel, rabies expression which labels the neurons were visualized in red channel and Nissl stained cell bodies were visualized in blue channel. Hence to segment the neurons, only the red channel of microscopic image was used. As the red channel was saturated, we observed that a single threshold of $255$ to segment the neurons gave reasonable results (Fig \ref{fig:neuronSegmentation}). To remove any noise which might be introduced, pixel clusters which were small in size (less than 5 pixels) were discarded. The remaining clusters represented the segmented neuronal cell bodies. Although most clusters contained single neurons, there were some clusters with multiple neuronal cell bodies. Hence, to resolve any ambiguity during counting of neurons for such clusters, we first computed the area of all the clusters and their median value was used to approximate the size of the neuronal cell body. We then took the ratio of the size of each cluster and the size of the neuronal cell body computed earlier and rounded it to the next nearest integer. The location of the neuronal cell body was represented by the centroid of the clusters. Using the annotations from ARA maps, location of each cluster and the number of neurons in each cluster, we then computed the number of neurons in all the 20 different anatomical regions for all the 51 registered microscopic slices (Table \ref{tab:neuronProfiling}).

\vspace{-0.1 cm}
\section{Discussion}
\label{discussion}

\begin{figure}[!b]
\centering
\includegraphics[width=0.49\textwidth]{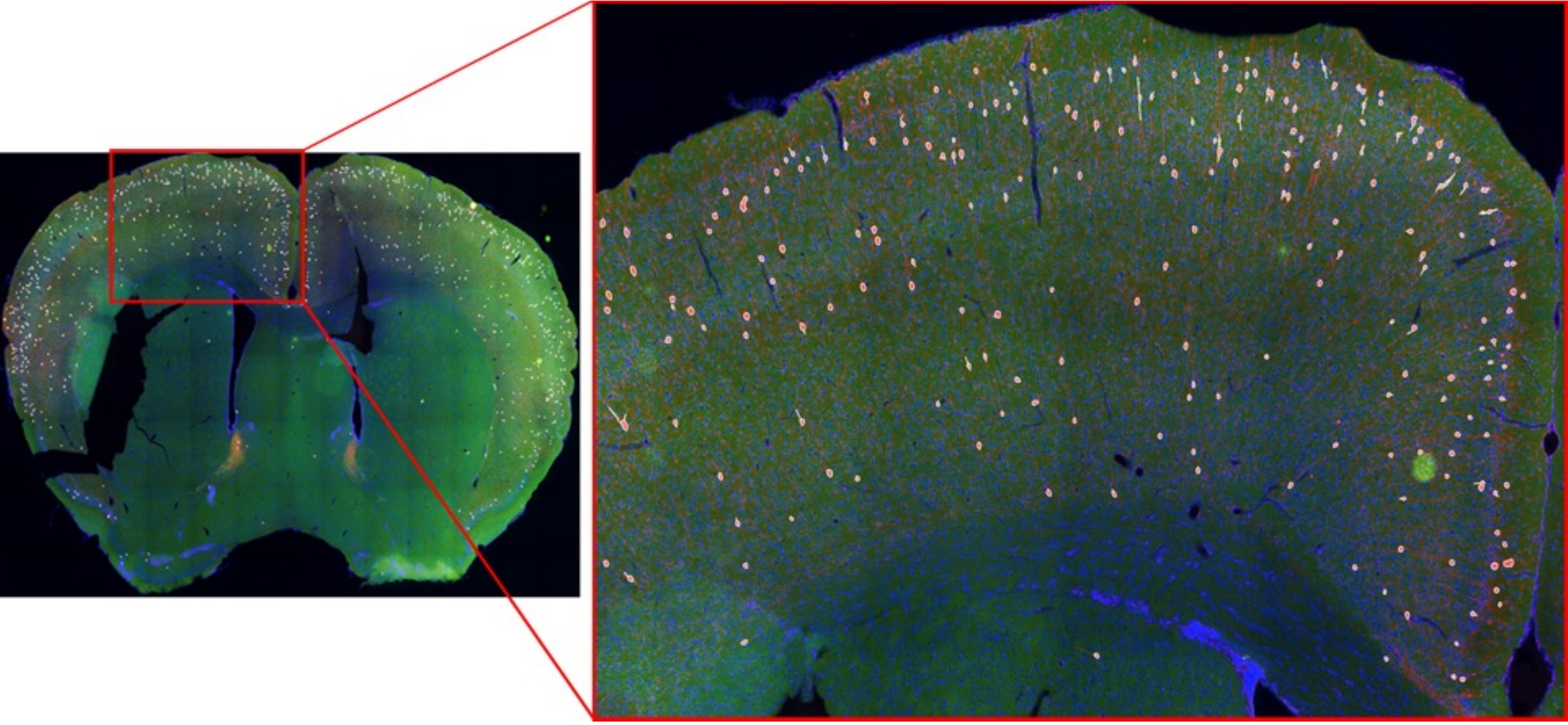}
\caption{\textbf{Neuron segmentation on microscope image:} Neuron segmentation was performed on a microscopic slice image where the segmented neuronal cell bodies (red) are surrounded by white contour. (please zoom in for details)}
\label{fig:neuronSegmentation}
\end{figure}

\begin{table}[!t]
\caption{Number of neurons in 20 different anatomical regions from 51 registered microscopic slices.}
\centering
\begin{tabular}{| R | N | R | N | }
\hline
\textbf{Anatomical Region} & \textbf{\# of Neurons} & \textbf{Anatomical Region} & \textbf{\# of Neurons}\\
\hline
\smallskip
Isocortex & 50267 & Olfactory areas & 6045 \\
\hline
\smallskip
Striatum Amygdalar & 1862 & Nucleus Accumbens & 1668 \\
\hline
\smallskip
Cortical Sub-Plate & 1386 & Hypothalamus & 1829 \\
\hline
\smallskip
Hippocampal Region & 618 & Olfactory Tubercle & 444 \\
\hline
\smallskip
Retrohippocampal & 315 & Caudoputamen & 228 \\
\hline
\smallskip
Lateral Septal Complex & 170 & Thalamus & 76 \\
\hline
\smallskip
Epithalamus & 22 & Fundus of Striatum & 32 \\
\hline
\smallskip
Midbrain & 16 & Pallidum & 0 \\
\hline
\smallskip
Pons & 0 & Medulla & 0 \\
\hline
\smallskip
Cerebellar Nuclie & 0 & Cerebellar Cortex & 0 \\
\hline
\end{tabular}
\label{tab:neuronProfiling}
\end{table}

Histological analysis is still the gold-standard for the accurate description of neuroanatomy and for tissue characterization of the mouse brain. Since our goal is to assist in creating a database by bringing together all microscopic slices from different histological preparations to a common anatomical framework (like ARA maps) and also gather various statistics about neuron densities in different regions and axonal projections of the mouse brain, it is vital that we achieve an accurate and robust registration. 

\noindent
\textbf{Generality:} In this work, we propose a novel feature based non-linear registration pipeline for automatic and robust alignment of high-resolution mouse brain microscopic slice images even with histological artifacts (tissue loss, tears, and deformation) to annotated atlas slice images from Allen Reference Atlas. We do this by aligning both the outer and inner contours of the microscopic and the annotated atlas slice images. As our method does not use the Nissl intensity slices of the atlas, we can register generic and typical cases of those microscopic slices that have a different intensity profile from that of the atlas, without using any intermediate atlas. This is significant as our approach gives the freedom to register microscopic slices from many sample preparation techniques and protocols, using a variety of imaging modality and of any scale with the ARA maps and bring them to a common anatomical reference framework. Another advantage of our method is that it can handle slice-specific histological artifacts such as tissue tears and tissue loss, which are very common in slices produced from conventional techniques. Both these benefits allow our algorithm to align more brain datasets with diverse profiles to ARA maps for more thorough and extensive connectome studies. 

\noindent
\textbf{Area/Volume analysis:} Although previous approaches \cite{Carson:2010aa,Ng:2007aa} map the atlas slices onto the microscopic slices, we mapped the microscopic slices onto the atlas slices for the ease of 3D visualization and neuron cell counting. During the mapping, obviously, area or volume of the input slices are changed. Once the mapping is done, regions can be demarcated in the original microscope slice. Inverting the mapping will take the region boundaries to the original tissue space, allowing for accurate area/volume measurements to detect abnormal growth of each region. Further after mapping, just like other approaches, we can store any image data (neurons, pixel intensity or other features) into the database for subsequent querying and analysis. 

\noindent
\textbf{Scalability:} Our robust damage detection algorithm can not only automatically detect slices which have artifacts, but also accurately locate the damaged regions in a slice without using any information from neighbouring slices. This makes our algorithm easily scalable to handle very large datasets without imposing any restrictions to the conventional neuroanatomical procedures. Further, our method can successfully detect multiple artifacts that may be present in the microscopic slices (see supplementary document). This enables and facilitates extremely thin sectioning of the mouse brain tissue, which is necessary for an accurate 3D mouse brain reconstruction. 

\noindent
\textbf{Advantages of 3D virtual mouse brain model:} During our work, we also performed surface reconstruction of 132 coronal annotated slices from ARA maps to generate a 3D virtual mouse brain model with 20 different anatomical regions. Such models, apart from assisting in visualizing the spatial location and orientation of the microscopic slices (Fig \ref{fig:pipeline}) and neurons (Fig \ref{fig:teaser}), have several other advantages. First, since the Allen Reference Atlas consists only of 2D slices at specific orientations and slicing intervals, a 3D virtual mouse brain model enables virtual slicing at any angle with arbitrary slicing intervals. This ensures that there is \emph{always} a corresponding annotated atlas slice for any given microscopic slice. Second, such models play a vital role in studying the connectome or the wiring diagram of the mouse brain \cite{Lichtman:2011aa}, computing the density and distribution of neurons \cite{Herculano-Houzel:2009aa} and analyzing the common gene expression patterns \cite{Ng:2007aa, Carson:2010aa}. Last, having a surface model like this, which is free from intersections, could be used as a precursor for building 3D sub-divisional based atlas \cite{Ju:2010aa} for faster multi-resolution querying. Although the current virtual mouse brain model has 20 regions, we are in the process of reconstructing close to 150 distinct anatomical regions.

\vspace{-0.1 cm}
\section{Limitations and Future Work}
\label{sec:futureWork}

Despite our registration algorithm being robust to tissue tears and tissue loss, there are still some extreme histological artifacts that are difficult to handle even with our algorithm. For example, tears that result in multiple component of tissues cannot be handled by our method. Such tissues are very difficult to mount since multiple components have to be accurately placed in their original positions onto the glass slide. In such cases, although our algorithm can detect multiple components, it cannot process such slices any further. Our algorithm also does not handle folding of the tissue or overlap of the adjacent tissue regions. For such artifacts a more complicated or semi-automatic approach might be helpful. Another extreme deformation present usually in the bottom (posterior) slices is the relative displacement of left and right lobes of the mouse brain tissue. For such slices, a combination of our method and a sub-region (block) registration might be more helpful \cite{Dauguet:2007aa}. Further, there might be some microscopic slices which do not have prominent features inside the tissue region, making our feature based registration technique ineffective. For such slices, a combination of our feature based and intensity based registration approaches could be used to accurately align both the interior and exterior regions of the microscopic slice. 

As one of the application of our registration pipeline, although we performed 2D segmentation of neuronal cell bodies by thresholding on image intensity, this could be vastly improved using more sophisticated approaches such as neural networks along with shape based classifiers or even performing 3D segmentation using active contours. Another possible future research direction could be to automatically compute the slicing angle of a microscopic slice in the virtual 3D mouse brain model. This would enable to register those microscopic slices for which the slicing angle is not known. 

Neural circuit mapping based on conventionally processed brain sections is riddled with histological artifacts and also is the most common form of data used in almost all neuroanatomical laboratories. Registration, analysis and visualization that we propose for such data, opens up endless possibilities of new research directions. 

\vspace{-0.1 cm}
\section*{Acknowledgments}
The authors would like to thank Dr Hong-Wei Dong for providing the mouse brain atlas images. We would also like to thank Xiaoxiao Lin and Yanjun Sun for sample preparation, imaging and collection of histological slice data. This work was supported in part by NIH grants (R01MH105427 and R01NS078434).

\end{document}